\documentclass[aps,pra,twocolumn,superscriptaddress,nopacs,letterpaper,longbibliography]{revtex4-1}

\usepackage[utf8]{inputenc}
\usepackage{amsmath}
\usepackage{color,graphicx}
\usepackage[pdfstartview=FitH,colorlinks=true,linkcolor=blue,citecolor=blue,urlcolor=blue]{hyperref}
\newcommand{\1}{\uparrow}
\newcommand{\2}{\downarrow}
\usepackage{lmodern}

\begin{document}

\title{Collective Excitations of a One-Dimensional Quantum Droplet}

\author{Marek Tylutki}
\email{marek.tylutki@pw.edu.pl}
\affiliation{Faculty of Physics, Warsaw University of Technology, Ulica Koszykowa 75, PL-00662 Warsaw, Poland}
\affiliation{Department of Applied Physics, Aalto University, FI-00076 Aalto, Finland}
\author{Grigori E. Astrakharchik}
\affiliation{Departament de F\'{\i}sica, Universitat Polit\`{e}cnica de Catalunya, E-08034 Barcelona, Spain}
\author{Boris A. Malomed}
\affiliation{Department of Physical Electronics, School of Electrical Engineering, Faculty of Engineering and The Center for Light-Matter Interaction, Tel Aviv University, Tel Aviv 69978, Israel}
\affiliation{Instituto de Alta Investigación, Universidad de Tarapacá, Casilla 7D, Arica, Chile}
\author{Dmitry S. Petrov}
\affiliation{Universit\'e Paris-Saclay, CNRS, LPTMS, 91405, Orsay, France}

\date{\today}

\begin{abstract}
We calculate the excitation spectrum of a one-dimensional self-bound quantum droplet in a two-component bosonic mixture described by the Gross-Pitaevskii equation (GPE) with cubic and quadratic nonlinearities. The cubic term originates from the mean-field energy of the mixture proportional to the effective coupling constant $\delta g$, whereas the quadratic nonlinearity corresponds to the attractive beyond-mean-field contribution. The droplet properties are governed by a control parameter $\gamma\propto \delta g N^{2/3}$, where $N$ is the particle number. For large $\gamma>0$ the droplet features the flat-top shape with the discrete part of its spectrum consisting of plane-wave Bogoliubov phonons propagating through the flat-density bulk and reflected by edges of the droplet. With decreasing $\gamma$ these modes cross into the continuum, sequentially crossing the particle-emission threshold at specific critical values. A notable exception is the breathing mode which we find to be always bound. The balance point $\gamma = 0$ provides implementation of a system governed by the GPE with an unusual quadratic nonlinearity. This case is characterized by the ratio of the breathing-mode frequency to the particle-emission threshold equal to 0.8904. As $\gamma$ tends to $-\infty$ this ratio tends to 1 and the droplet transforms into the soliton solution of the integrable cubic GPE. 
\end{abstract}

\maketitle

The mean-field interaction in a mixture of two bosonic superfluids can be fine tuned to small values and can thus become comparable to the beyond-mean-field (BMF) energy correction, originating from the celebrated work by Lee, Huang, and Yang~\cite{LHY}. In this case, quantum many-body effects crucially manifest themselves in spite of the fact that the system is in the weakly interacting regime. In particular, in three dimensions, the competition between the slightly attractive mean-field term and repulsive beyond-mean-field one can lead to the formation of self-bound droplet states as predicted theoretically and demonstrated experimentally~\cite{Petrov2015,Cabrera2018,Cheiney2018,Semeghini2018,Ferioli2019,DErrico2019}. We should mention here related studies of self-bound Bose-Fermi mixtures~\cite{Rakshit2019,Rakshit2019a} and a very quickly developing field of dipolar droplets~\cite{Ferrier-Barbut2016,Chomaz2016,Ferrier-Barbut2018,Tanzi2019,Chomaz2019,Bottcher2019,Tanzi2019Modes,Guo2019}, also stabilized by BMF effects \cite{Wachtler2016,Wachtler2016a,Bisset2016}.

\begin{figure}
\includegraphics[clip = true, width = .99\columnwidth]{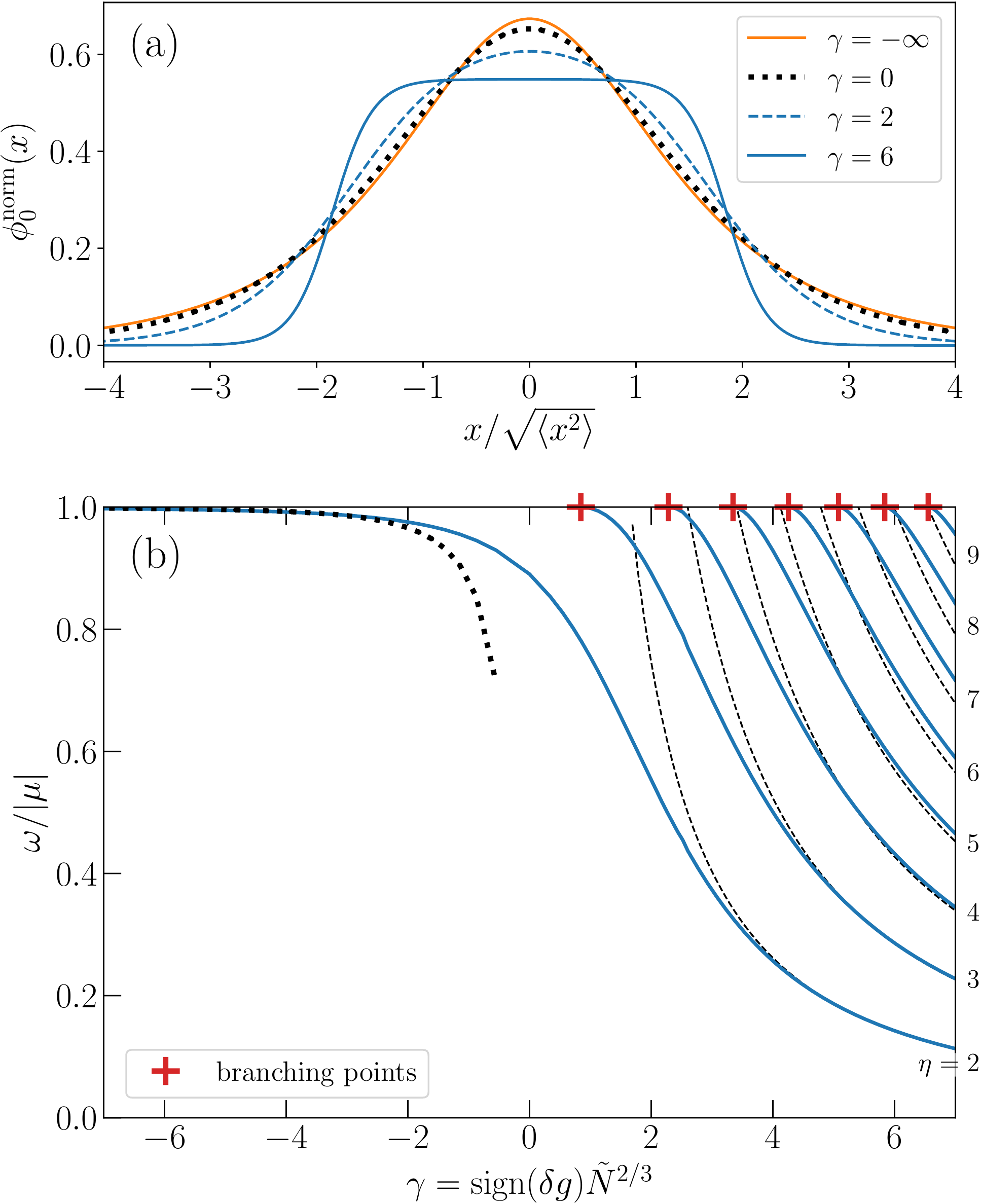}
\caption{(a) The droplet wave functions for different values of the control parameter $\gamma = -\infty$, $0$, $2$, and $6$, ranging from the soliton to flat-top shapes. For better comparison we normalize the corresponding density profiles to 1 and rescale $x$ so that they have the same rms widths. (b) The ratio of discrete Bogoliubov frequencies $\omega_\eta$ to the particle-emission threshold $-\mu$ as a function of $\gamma={\rm sign}(\delta g)\tilde{N}^{2/3}$. Red crosses indicate branching points where the discrete modes cross the particle-emission threshold and enter the continuum of excitations. The dashed and dotted curves represent asymptotic approximations~(\ref{eq.omega_eta}) and (\ref{eq.omega_2}), correspondingly.}
\label{fig.bog}
\end{figure}

The BMF contribution, being produced by zero-point energies of all Bogoliubov modes, strongly depends on the density of states and, thus, on the dimensionality of the system. For this reason, low-dimensional droplets, especially one-dimensional (1D) ones, fundamentally differ from their three-dimensional (3D) counterparts~\cite{Petrov2016} offering, in particular, significant practical advantages in terms of stability. Quite generally, BMF effects in 1D systems may be enhanced by decreasing the density without compromising the system's lifetime. Another feature contrasting with the 3D case is that in one dimension the BMF energy correction is negative and the self-binding property is manifest for any particle number and for any sign of the mean-field term. In this context, an important practical question is how to distinguish droplets from quasi-1D bright solitons observed in attractive single-component quantum gases \cite{Khaykovich2002,Strecker2002,Marchant2013,Medley2014,Nguyen2014,Lepoutre2016,DiCarli2019}. Note that the 1D quadratic-cubic Gross-Pitaevskii equation (GPE) describing the droplets provides the full analytical account of their static properties, such as shapes and energies\cite{Hayata1995,Triki2017,Mithun2020}. However, differences in the shapes of solitons and droplets become apparent only when the latter are locate deeply in the flat-top regime [see Fig.~\ref{fig.bog}(a)]. The same can also be said about their dynamics, which is nonintegrable in the droplet case; in particular, the inelastic aspects of the droplet-droplet collision are more visible at higher energies and for droplets with a significant flat-top region~\cite{Astrakharchik2018}. Another interesting, experimentally measurable, but so far unexplored dynamical property of 1D droplets is the spectrum of small-amplitude excitations. In this respect, the droplets are expected to feature qualitative differences in comparison to 1D solitons, which support no small-amplitude collective modes, but solely the continuum spectrum (see, for example, \cite{Kaup1990,Pelinovsky1998}), and in comparison to 3D droplets, with their quite peculiar behavior of bulk and surface collective modes~\cite{Petrov2015}.

In this Rapid Communication we investigate the excitation spectrum of a quantum droplet formed in a weakly interacting 1D mass-balanced binary mixture with competing mean-field interactions tuned to a weak overall repulsion or attraction ($\propto \delta g$). In this case, integrating out the relative motion and including the corresponding BMF correction, one arrives at a scalar GPE with cubic (mean-field) and attractive quadratic (BMF) nonlinearities. This GPE can be cast in a dimensionless form controlled only by a dimensionless parameter $\gamma\propto \delta g N^{2/3}$, which is exactly defined in Eq.~(\ref{eq.gamma}). It determines the droplet's shape and its excitation frequencies, which we calculate from the linearized Bogoliubov-de Gennes equations and plot in Fig.~\ref{fig.bog}(b). We find that, for large $\gamma$, flat-top droplets with a large bulk region support plane-wave phonons and behave similarly to constant-density elastic media with free ends. As $\gamma$ decreases, the droplet's bulk region shrinks and the phonons get pushed above the particle-emission threshold, in which those with smaller wave lengths get to the continuum faster; see Fig.~\ref{fig.bog}(b). Eventually, the breathing oscillation remains the only collective discrete mode supported by the droplet. For $\gamma=0$, the cubic nonlinearity disappears and our system is described by a GPE with a rather unusual quadratic-only nonlinearity. In this case, the ratio of the breathing-mode frequency to the particle-emission threshold equals $\omega_2/(-\mu) \approx 0.8904$, clearly distinguishing our droplet from the soliton of the cubic GPE. This ratio tends to +1 for large negative $\gamma$, when the droplet crosses over to the soliton.

We consider a binary mass-balanced Bose-Bose mixture with two components labeled by $\sigma = \{\uparrow,\downarrow\}$, which is characterized by the set of three coupling constants $g_{\sigma\sigma'}=g_{\sigma'\sigma}$. In the vicinity of the mean-field collapse point (i.e., for small $\delta g = g_{\1\2} + \sqrt{g_{\1\1} g_{\2\2}} \ll g_{\1\1}\sim g_{\2\2}$) the energy density of the homogeneous mixture reads~\cite{Petrov2016}
\begin{eqnarray}\label{eq.EnergyDensity}
E&\approx &\frac{(g_{\1\1}^{1/2}n_\1-g_{\2\2}^{1/2}n_\2)^2}{2}+\frac{\sqrt{g_{\1\1} g_{\2\2}}\delta g(g_{\2\2}^{1/2}n_\1+g_{\1\1}^{1/2}n_\2)^2}{(g_{\1\1}+g_{\2\2})^2}\nonumber \\
&&\hspace{-0.cm}-\frac{2\sqrt{m}}{3\pi\hbar}(g_{\1\1}n_\1+g_{\2\2}n_\2)^{3/2} ~,
\end{eqnarray}
where $n_\1$ and $n_\2$ are the densities of the components and $m$ is the mass of the atoms. The first and second terms on the right-hand side of Eq.~(\ref{eq.EnergyDensity}) correspond to the mean-field contribution and the last one is the leading-order BMF correction accounting for quantum many-body effects. The expansion is valid in the weakly interacting regime $|g_{\sigma\sigma'}|/n\ll 1$ and, as we have mentioned, it requires $|\delta g|$ to be small. Under these conditions, it was demonstrated~\cite{Petrov2015,Petrov2016} that the mixture can exist in vacuum as a droplet. The minimization of the dominant first term in the right-hand side of Eq.~(\ref{eq.EnergyDensity}) forces the components of the mixture to follow the ratio $n_\1/n_\2=\sqrt{g_{\2\2}/g_{\1\1}}$, which is also true in the inhomogeneous case for sufficiently smooth temporal and spatial variations of the total density. The mixture thus reduces to an effectively single-component Bose liquid parametrized by $\psi(x)$, related to the individual component wave functions by $\psi_{\sigma}(x)=g_{\bar{\sigma}\bar{\sigma}}^{1/4}\psi(x)/\sqrt{\sqrt{g_{\sigma\sigma}}+\sqrt{g_{\bar{\sigma}\bar{\sigma}}}}$, where $\bar{\1}=\2$ and $\bar{\2}=\1$. Accordingly, the Gross-Pitaevskii energy functional of the two fields $\psi_{\1}(x)$ and $\psi_{\2}(x)$, the local part of which is given by Eq.~(\ref{eq.EnergyDensity}), reduces to a functional of $\psi(x)$ and leads to the GPE,
\begin{eqnarray}\label{eq.GPgeneralGP1D}
i\hbar\partial_t \psi &=& -\frac{\hbar^2}{2m}\partial_x^2\psi+\frac{2\sqrt{g_{\1\1}g_{\2\2}}\delta g}{(\sqrt{g_{\1\1}}+\sqrt{g_{\2\2}})^2}|\psi|^2\psi \nonumber \\
&&-\frac{\sqrt{m}}{\pi\hbar}(g_{\1\1}g_{\2\2})^{3/4}|\psi|\psi ~,
\end{eqnarray}
with the total density given by $n(x)=|\psi(x)|^2$ and total atom number $N=\int_{\infty}^\infty |\psi(x)|^2dx$. Introducing the healing length
\begin{equation}\label{eq.HealingLength}
\xi =\frac{\pi\hbar^2}{m}\frac{\sqrt{2|\delta g|}}{\sqrt{g_{\1\1}g_{\2\2}}(\sqrt{g_{\1\1}}+\sqrt{g_{\2\2}})},
\end{equation} 
and rescaling coordinate $x=\xi\tilde{x}$, time $t=(\hbar/m\xi^2)\tilde{t}$, and the wave function  
\begin{equation}\label{eq.psiRescaling}
\psi=\frac{(\sqrt{g_{\1\1}}+\sqrt{g_{\2\2}})^{3/2}}{\sqrt{\pi \xi} (2|\delta g|)^{3/4}}\phi,
\end{equation}
Eq.~(\ref{eq.GPgeneralGP1D}) acquires the dimensionless form
\begin{equation}
i \partial_{\tilde{t}} \phi(\tilde{x}, \tilde{t}) = [-\partial_{\tilde{x}}^2/2 + {\rm sign}(\delta g)|\phi(\tilde{x}, \tilde{t})|^2 - |\phi(\tilde{x}, \tilde{t})|] \phi(\tilde{x,}\tilde{t})
\label{eq.gp}
\end{equation}
with the normalization condition $\int_{-\infty}^\infty |\phi(\tilde{x})|^2d\tilde{x}=\tilde{N}=N\pi(2|\delta g|)^{3/2}/(\sqrt{g_{\1\1}}+\sqrt{g_{\2\2}})^3$. Together with the sign of $\delta g$, the rescaled atom number $\tilde{N}$ provides a parametrization of the system. Both parameters can be combined into the single dimensionless coupling constant
\begin{equation}
\gamma={\rm sign}(\delta g)\tilde{N}^{2/3}=2(\pi N)^{2/3}\delta g/(\sqrt{g_{\1\1}}+\sqrt{g_{\2\2}})^2.
\label{eq.gamma}
\end{equation}
From now on we adopt the rescaled units and omit tildes over $x$, $t$ and $N$, measuring energy in units of $\hbar^2/m\xi^2$, frequencies in units of $\hbar/m\xi^2$, etc. 

The ground-state wave function $\phi_0$ is a stationary solution of Eq.~(\ref{eq.gp}), $\phi(x, t) = \phi_0(x) e^{-i\mu t}$, with the always attractive BMF term allowing for a self-localized solution in the form of a droplet, irrespective of the sign of $\delta g$. The ground-state solution is~\cite{Petrov2016,Mithun2020}
\begin{equation}
\phi_0(x) = \frac{ \sqrt{n_0}\, \mu / \mu_0}{ 1 + \sqrt{1 - {\rm sign}(\delta g)\mu / \mu_0}\, \cosh(\sqrt{-2 \mu} x) }
\label{eq.phi0}
\end{equation}
with the relation between the chemical potential and the rescaled particle number
\begin{equation}
N_{\delta g>0}=n_0\sqrt{-\frac{2}{\mu_0}}\left[\ln\frac{1+\sqrt{\mu/\mu_0}}{\sqrt{1-\mu/\mu_0}} - \sqrt{\frac{\mu}{\mu_0}}\right]
\label{eq.Nvsmu}
\end{equation}
\begin{equation}
N_{\delta g<0}=n_0\sqrt{-\frac{2}{\mu_0}}\left(\sqrt{\frac{\mu}{\mu_0}}-\arctan\sqrt{\frac{\mu}{\mu_0}}\right)
\label{eq.NvsmuNeg}
\end{equation}
depending on the sign of $\delta g$. In Eqs.~(\ref{eq.Nvsmu}) and (\ref{eq.NvsmuNeg}) $\mu_0 = -2 / 9$ and $n_0=4/9$ are, respectively, the chemical potential and the saturation density for a uniform liquid at $\delta g>0$. 

We distinguish three characteristic regimes of the droplet's behavior [see Fig.~\ref{fig.bog}(a) for respective droplet profiles]. The first is the flat-top case, with $\delta g>0$ and $N\gg 1$ (i.e., $\gamma\gg 1$). Note that in 1D the surface of the droplet reduces to two edges, and therefore the surface tension does not significantly affect the bulk density. Namely, for large $N$ the droplet's bulk density deviates from $n_0=4/9$ by an exponentially small correction $n \approx n_0 [1 + 4 \exp(-1 - 3 N / 4)]$ and the chemical potential is $\mu\approx \mu_0[1-4\exp(-2-3N/2)]$ \cite{Astrakharchik2018}. Therefore, the right matter-vacuum interface of the droplet in this regime, with exponential accuracy, reduces to the kink structure,
\begin{equation}
\phi_{\mathrm{kink}}(x)=\frac{2/3}{1+\exp(2x/3-1-L/3)},  \label{kink}
\end{equation}
which is an exact solution of Eq.~(\ref{eq.gp}) with $\mu=\mu_0$. For convenience, we have set the horizontal shift in Eq.~(\ref{kink}) such that the center of mass of the kink's density is located at $x=L/2$, where $L=N/n_0$ is the length of the droplet. The left edge is represented by the respective antikink $\phi _{\mathrm{kink}}(-x)$.

The second limit corresponds to small $\gamma$, i.e., small $N$ and $|\mu|$. In this case, the cubic nonlinear term in Eq.~(\ref{eq.gp}) can be neglected and we arrive at a GPE with a rather unusual quadratic-only nonlinearity, which gives rise to the droplet wave function of the Korteweg - de Vries type $1/\cosh^2(\sqrt{-\mu/2}\, x)$.  

The third limit corresponds to negative $\delta g$, large $N$, and the chemical potential diverging as $\mu\propto -N^2$ [see Eq.~(\ref{eq.NvsmuNeg})]. In this regime, the quadratic nonlinearity can be neglected, and Eq.~(\ref{eq.gp}) becoming the integrable GPE with the cubic nonlinearity characterized by the bright soliton solution 
\begin{equation}\label{soliton}
\phi_{\rm s}(x)= \sqrt{n_0\mu/\mu_0}/\cosh (\sqrt{-2\mu}\, x). 
\end{equation}

Small fluctuations of the droplet can be calculated by linearizing GPE~(\ref{eq.gp}) around the ground state given by Eq.~(\ref{eq.phi0}). Namely, writing 
\begin{equation}
\phi(x, t) = 
e^{-i \mu t}\!\left\{\!
\phi_0(x) + \sum_\eta \left[ u_\eta(x) e^{-i \omega_\eta t} + v^*_\eta(x) e^{i \omega_\eta t} \right]\!\right\}
\end{equation}
and expanding Eq.~(\ref{eq.gp}) to the first order in $u$ and $v$, we obtain the Bogoliubov-de Gennes equations,
\begin{equation}
\begin{bmatrix}
\mathcal{T} - \omega_\eta & {\rm sign}(\delta g)\phi_0^2 - \frac12 \phi_0 \\
{\rm sign}(\delta g)\phi_0^2 - \frac12 \phi_0 & \mathcal{T} + \omega_\eta \\
\end{bmatrix} \begin{bmatrix} u_\eta(x)\\ v_\eta(x) \\ \end{bmatrix} = 0 ~,
\label{eq.bog}
\end{equation}
where the operator is $\mathcal{T}=-\partial_x^2/2-\mu +2{\rm sign}(\delta g)\phi_0^2-3\phi_0/2$ and we use the fact that $\phi_0$ is real. We solve Eqs.~(\ref{eq.bog}) numerically finding discrete eigenfrequencies $\omega_\eta$, labeled by integer $\eta$. The value $\eta=0$ stands for the zero-frequency mode, proportional to the droplet wave function itself, representing an infinitesimal phase shift, $\phi_0 \to \phi_0(x) + i \alpha \phi_0(x)$. The excitation with $\eta=1$ corresponds to the center-of-mass displacement of the droplet which has a vanishing frequency. The lowest nontrivial collective mode in our setup is the breathing mode with $\eta=2$. 

Ratios of the mode frequencies to the particle-emission threshold, $-\omega_\eta/\mu$ are shown in Fig.~\ref{fig.bog}(b) as functions of parameter $\gamma$ defined in Eq.~(\ref{eq.gamma}). This spectrum is the main result of this Rapid Communication. We find that the breathing mode always stays below the particle-emission threshold, whereas all other modes with $\eta \geq 3$ eventually cross it, following the decrease of $\gamma$. Near the crossings the corresponding mode is characterized by a large probability of finding a particle (nonvanishing $u_\eta$) outside of the droplet. In this case, one may treat the droplet as a potential well for atoms, the depth of which changes linearly with $N$ close to the crossing point. The corresponding particle-droplet binding energy then follows the usual 1D threshold law $-\mu-\omega_\eta\propto (N-N_\eta)^2$, consistent with our numerical results. The threshold values for a few lowest modes are $N_3 \approx 0.774$, $N_4 \approx 3.453$, $N_5 \approx 6.119$, $N_6 \approx 8.783$, $N_7 \approx 11.447$.

We now address the structure of the modes in the flat-top limit, $\gamma\gg 1$, in the regime $\omega_\eta \ll -\mu$. In this case, Eqs.~(\ref{eq.bog}) can be diagonalized in terms of plane waves in the bulk of the droplet where, as said above, one might set $\phi_0=2/3$ and $\mu=-2/9$. Since $\phi_0(x)=\phi_0(-x)$, the solutions are then either even (cos) or odd (sin) combinations of plane waves
\begin{equation}\label{eq.bulksolution}
\begin{bmatrix} u_\eta(x)\\ v_\eta(x) \\ \end{bmatrix}    \propto \begin{bmatrix} 1/9\\ \omega_\eta-\sqrt{\omega_\eta^2+1/81} \\ \end{bmatrix}(e^{ik_\eta x}\pm e^{-ik_\eta x}),
\end{equation}
where $k_\eta^2/2=\sqrt{\omega_\eta^2+1/81}-1/9$. To find the eigenfrequency $\omega_\eta$ one should match (\ref{eq.bulksolution}) to a solution of Eqs.~(\ref{eq.bog}) around the right edge of the droplet, where these equations take the form of
\begin{subequations}
\begin{eqnarray}
\hat{\xi}_+ f^{+}_\eta(x)&=\omega_\eta f^{-}_\eta(x),\label{fplus}\\
\hat{\xi}_- f^{-}_\eta(x)&=\omega_\eta f^{+}_\eta(x),\label{fminus}
\end{eqnarray}
\end{subequations}
where $f^{\pm}_\eta=u_\eta(x) \pm v_\eta(x)$, and the operators are
\begin{equation}
\hat{\xi}_\pm = -\partial_x^2/2 - \mu_0+(2\pm 1)\phi_{\mathrm{kink}}^2(x)-(3/2\pm 1/2) \phi_{\mathrm{kink}}(x)
\end{equation}
and we have neglected the exponentially small deviation of $\phi_0$ from $\phi_{\mathrm{kink}}$ and $\mu$ from $\mu_0$. For $\omega_\eta = 0$ Eqs.~(\ref{fminus}) and (\ref{fplus}) decouple and are solved by arbitrary combinations of $f^{-}(x)\propto \phi_{\mathrm{kink}}(x)$ and $f^{+}(x)\propto \partial_x \phi_{\mathrm{kink}}(x)=- 1/[9\cosh^2(x/3-1/2-L/6)]$, which correspond, respectively, to uniform phase rotation and translation of the droplet's edge~\cite{Birnbaum2008}. For small finite $\omega_\eta$ one can iterate Eqs.~(\ref{fplus}-\ref{fminus}) obtaining their solution in powers of $\omega_\eta$. Neglecting the phase rotation we choose $f_\eta^+=\partial_x \phi_{\mathrm{kink}}$, $f_{\eta}^-=0$ as the zero-order solution which we substitute in the right-hand side of Eqs.~(\ref{fplus}-\ref{fminus}). The first iteration gives
\begin{equation}\label{eq.family}
\begin{bmatrix}
f^{+}_\eta \\
f^{-}_\eta
\end{bmatrix}
= \begin{bmatrix}
\partial_x \phi_{\mathrm{kink}}(x) \\
0
\end{bmatrix}+\omega_\eta
\begin{bmatrix}
0 \\
(a-x)\phi_{\mathrm{kink}}(x)
\end{bmatrix},
\end{equation}
where $a$ is an arbitary constant. This constant is determined in the next iteration by noting that the equation $\hat{\xi}_+ f^{+}_\eta(x)=\omega_\eta^2 (a-x)\phi_{\mathrm{kink}}(x)$ can be solved only when its right-hand side is orthogonal to $\partial_x \phi_{\mathrm{kink}}(x)$ since the latter corresponds to a discrete eigenstate of $\hat{\xi}_+$ with a vanishing eigenvalue. The condition $\int (a-x)\phi_{\rm kink}(x)\partial_x \phi_{\mathrm{kink}}(x) dx=0$ gives $a=L/2$. Matching expression~(\ref{eq.family}) with $a=L/2$ to plane waves (\ref{eq.bulksolution}) yields the spectrum
\begin{equation}
\omega_\eta \approx 4 \pi (\eta - 1) / (27 N)
\label{eq.omega_eta}
\end{equation}
shown in Fig.~\ref{fig.bog}(a) by dashed curves. The fact that the (extrapolated) node of the phonon field is located at $x=L/2$ is consistent with the fact that the edge of the droplet is free, i.e., it experiences no compression and no gradient of the velocity field.

As we approach the opposite (soliton) limit $\gamma\rightarrow -\infty$, we observe that the breathing mode frequency tends to, but never crosses, the particle-emission threshold. This is not a numerical artifact. The asymptotic behavior of this mode can be understood from the following perturbative procedure. Retaining only the leading-order terms in $1/\mu$ in Eqs.~(\ref{eq.bog}) [equivalent to the formal substitution $\phi_0^2\rightarrow \phi_{\rm s}^2$ and $\phi_0\rightarrow 0$] the resulting equations are solved by $\omega_2=-\mu$, $u(x)=\tanh^2(\sqrt{-2\mu}\, x)$, and $v(x)=-1/\cosh^2(\sqrt{-2\mu}\, x)$. This solution corresponds to the $s$-wave scattering of an atom by a soliton at zero collision energy and is characterized by infinite scattering length, i.e., there is, effectively, no atom-soliton interaction. Using the first-order perturbation theory around this solution [using $\phi_0^2-\phi_s^2$ and $\phi_0=\phi_s$ as perturbations in Eqs.~(\ref{eq.bog})] we find that the droplet acts as a weakly attractive potential for the atom, characterized by the scattering length $6/\pi$, which gives a weakly bound state (the breathing mode) with the energy
\begin{equation}
\omega_2+\mu \approx -\frac{\pi^2}{72}\ll -\mu.
\label{eq.omega_2}    
\end{equation}
In Fig.~\ref{fig.bog}(b) this asymptote is shown with a dotted bold line. Lastly, we note that the existence and properties of internal modes in somewhat similar cubic-quintic GPE were considered in Ref.~\cite{Pelinovsky1998} and a perturbative treatment of the quintic term for this GPE was also performed in the context of quasi-1D bright solitons~\cite{Sinha2006}.   

In conclusion, we have obtained the complete frequency spectrum of the one-dimensional self-bound quantum droplet of the Bose-Bose mixture. The results can be used to characterize the droplet, measure its parameters, and distinguish it from the bright soliton. By manipulating interactions in the mixture, all collective excitations, except the breathing mode, can be pushed into the continuum thus offering a way to cool the droplet. The breathing mode, which we find to be always bound, opens the way to experimental realization of a robust intrinsic mode in self-trapped matter-wave states (see \cite{Pelinovsky1998} for the discussion). We note that our derivation assumes a purely one-dimensional mixture in free space in the weakly interacting regime. Deviations from these assumptions --- in particular, effects of an external trapping and 3D character of the system (cf.~\cite{Carr2002,Cheiney2018,DiCarli2019}) --- may become qualitatively important in the experimental situation. We leave these topics for future studies.

\bigskip \noindent
{\it Acknowledgements ---} 
MT was supported by the Polish National Science Center (NCN), Contract No. UMO-2017/26/E/ST3/00428 and, at the initial stage, by the Academy of Finland under Projects No. 307419, No. 303351, and No. 318987. 
GEA is grateful to Aalto University for hospitality through the Centre for Quantum Engineering and acknowledges funding from the Spanish MINECO (FIS2017-84114-C2-1-P).
The work of BAM is supported in part by the Israel Science Foundation through grant No. 1286/17. DSP acknowledges support from ANR Grant Droplets No. ANR-19-CE30-0003-02.


\begin{thebibliography}{36}%
\makeatletter
\providecommand \@ifxundefined [1]{%
 \@ifx{#1\undefined}
}%
\providecommand \@ifnum [1]{%
 \ifnum #1\expandafter \@firstoftwo
 \else \expandafter \@secondoftwo
 \fi
}%
\providecommand \@ifx [1]{%
 \ifx #1\expandafter \@firstoftwo
 \else \expandafter \@secondoftwo
 \fi
}%
\providecommand \natexlab [1]{#1}%
\providecommand \enquote  [1]{``#1''}%
\providecommand \bibnamefont  [1]{#1}%
\providecommand \bibfnamefont [1]{#1}%
\providecommand \citenamefont [1]{#1}%
\providecommand \href@noop [0]{\@secondoftwo}%
\providecommand \href [0]{\begingroup \@sanitize@url \@href}%
\providecommand \@href[1]{\@@startlink{#1}\@@href}%
\providecommand \@@href[1]{\endgroup#1\@@endlink}%
\providecommand \@sanitize@url [0]{\catcode `\\12\catcode `\$12\catcode
  `\&12\catcode `\#12\catcode `\^12\catcode `\_12\catcode `\%12\relax}%
\providecommand \@@startlink[1]{}%
\providecommand \@@endlink[0]{}%
\providecommand \url  [0]{\begingroup\@sanitize@url \@url }%
\providecommand \@url [1]{\endgroup\@href {#1}{\urlprefix }}%
\providecommand \urlprefix  [0]{URL }%
\providecommand \Eprint [0]{\href }%
\providecommand \doibase [0]{http://dx.doi.org/}%
\providecommand \selectlanguage [0]{\@gobble}%
\providecommand \bibinfo  [0]{\@secondoftwo}%
\providecommand \bibfield  [0]{\@secondoftwo}%
\providecommand \translation [1]{[#1]}%
\providecommand \BibitemOpen [0]{}%
\providecommand \bibitemStop [0]{}%
\providecommand \bibitemNoStop [0]{.\EOS\space}%
\providecommand \EOS [0]{\spacefactor3000\relax}%
\providecommand \BibitemShut  [1]{\csname bibitem#1\endcsname}%
\let\auto@bib@innerbib\@empty
\bibitem [{\citenamefont {Lee}\ \emph {et~al.}(1957)\citenamefont {Lee},
  \citenamefont {Huang},\ and\ \citenamefont {Yang}}]{LHY}%
  \BibitemOpen
  \bibfield  {author} {\bibinfo {author} {\bibfnamefont {T.~D.}\ \bibnamefont
  {Lee}}, \bibinfo {author} {\bibfnamefont {K.}~\bibnamefont {Huang}}, \ and\
  \bibinfo {author} {\bibfnamefont {C.~N.}\ \bibnamefont {Yang}},\ }\bibfield
  {title} {\enquote {\bibinfo {title} {Eigenvalues and eigenfunctions of a
  {Bose} system of hard spheres and its low-temperature properties},}\ }\href
  {\doibase 10.1103/PhysRev.106.1135} {\bibfield  {journal} {\bibinfo
  {journal} {Phys. Rev.}\ }\textbf {\bibinfo {volume} {106}},\ \bibinfo {pages}
  {1135} (\bibinfo {year} {1957})}\BibitemShut {NoStop}%
\bibitem [{\citenamefont {Petrov}(2015)}]{Petrov2015}%
  \BibitemOpen
  \bibfield  {author} {\bibinfo {author} {\bibfnamefont {D.~S.}\ \bibnamefont
  {Petrov}},\ }\bibfield  {title} {\enquote {\bibinfo {title} {Quantum
  mechanical stabilization of a collapsing {Bose-Bose} mixture},}\ }\href
  {\doibase 10.1103/PhysRevLett.115.155302} {\bibfield  {journal} {\bibinfo
  {journal} {Phys. Rev. Lett.}\ }\textbf {\bibinfo {volume} {115}},\ \bibinfo
  {pages} {155302} (\bibinfo {year} {2015})}\BibitemShut {NoStop}%
\bibitem [{\citenamefont {Cabrera}\ \emph {et~al.}(2018)\citenamefont
  {Cabrera}, \citenamefont {Tanzi}, \citenamefont {Sanz}, \citenamefont
  {Naylor}, \citenamefont {Thomas}, \citenamefont {Cheiney},\ and\
  \citenamefont {Tarruell}}]{Cabrera2018}%
  \BibitemOpen
  \bibfield  {author} {\bibinfo {author} {\bibfnamefont {C.~R.}\ \bibnamefont
  {Cabrera}}, \bibinfo {author} {\bibfnamefont {L.}~\bibnamefont {Tanzi}},
  \bibinfo {author} {\bibfnamefont {J.}~\bibnamefont {Sanz}}, \bibinfo {author}
  {\bibfnamefont {B.}~\bibnamefont {Naylor}}, \bibinfo {author} {\bibfnamefont
  {P.}~\bibnamefont {Thomas}}, \bibinfo {author} {\bibfnamefont
  {P.}~\bibnamefont {Cheiney}}, \ and\ \bibinfo {author} {\bibfnamefont
  {L.}~\bibnamefont {Tarruell}},\ }\bibfield  {title} {\enquote {\bibinfo
  {title} {Quantum liquid droplets in a mixture of {Bose-Einstein}
  condensates},}\ }\href {\doibase 10.1126/science.aao5686} {\bibfield
  {journal} {\bibinfo  {journal} {Science}\ }\textbf {\bibinfo {volume}
  {359}},\ \bibinfo {pages} {301--304} (\bibinfo {year} {2018})}\BibitemShut
  {NoStop}%
\bibitem [{\citenamefont {Cheiney}\ \emph {et~al.}(2018)\citenamefont
  {Cheiney}, \citenamefont {Cabrera}, \citenamefont {Sanz}, \citenamefont
  {Naylor}, \citenamefont {Tanzi},\ and\ \citenamefont
  {Tarruell}}]{Cheiney2018}%
  \BibitemOpen
  \bibfield  {author} {\bibinfo {author} {\bibfnamefont {P.}~\bibnamefont
  {Cheiney}}, \bibinfo {author} {\bibfnamefont {C.~R.}\ \bibnamefont
  {Cabrera}}, \bibinfo {author} {\bibfnamefont {J.}~\bibnamefont {Sanz}},
  \bibinfo {author} {\bibfnamefont {B.}~\bibnamefont {Naylor}}, \bibinfo
  {author} {\bibfnamefont {L.}~\bibnamefont {Tanzi}}, \ and\ \bibinfo {author}
  {\bibfnamefont {L.}~\bibnamefont {Tarruell}},\ }\bibfield  {title} {\enquote
  {\bibinfo {title} {Bright soliton to quantum droplet transition in a mixture
  of {Bose-Einstein} condensates},}\ }\href {\doibase
  10.1103/PhysRevLett.120.135301} {\bibfield  {journal} {\bibinfo  {journal}
  {Phys. Rev. Lett.}\ }\textbf {\bibinfo {volume} {120}},\ \bibinfo {pages}
  {135301} (\bibinfo {year} {2018})}\BibitemShut {NoStop}%
\bibitem [{\citenamefont {Semeghini}\ \emph {et~al.}(2018)\citenamefont
  {Semeghini}, \citenamefont {Ferioli}, \citenamefont {Masi}, \citenamefont
  {Mazzinghi}, \citenamefont {Wolswijk}, \citenamefont {Minardi}, \citenamefont
  {Modugno}, \citenamefont {Modugno}, \citenamefont {Inguscio},\ and\
  \citenamefont {Fattori}}]{Semeghini2018}%
  \BibitemOpen
  \bibfield  {author} {\bibinfo {author} {\bibfnamefont {G.}~\bibnamefont
  {Semeghini}}, \bibinfo {author} {\bibfnamefont {G.}~\bibnamefont {Ferioli}},
  \bibinfo {author} {\bibfnamefont {L.}~\bibnamefont {Masi}}, \bibinfo {author}
  {\bibfnamefont {C.}~\bibnamefont {Mazzinghi}}, \bibinfo {author}
  {\bibfnamefont {L.}~\bibnamefont {Wolswijk}}, \bibinfo {author}
  {\bibfnamefont {F.}~\bibnamefont {Minardi}}, \bibinfo {author} {\bibfnamefont
  {M.}~\bibnamefont {Modugno}}, \bibinfo {author} {\bibfnamefont
  {G.}~\bibnamefont {Modugno}}, \bibinfo {author} {\bibfnamefont
  {M.}~\bibnamefont {Inguscio}}, \ and\ \bibinfo {author} {\bibfnamefont
  {M.}~\bibnamefont {Fattori}},\ }\bibfield  {title} {\enquote {\bibinfo
  {title} {Self-bound quantum droplets in atomic mixtures},}\ }\href {\doibase
  10.1103/PhysRevLett.120.235301} {\bibfield  {journal} {\bibinfo  {journal}
  {Phys. Rev. Lett.}\ }\textbf {\bibinfo {volume} {120}},\ \bibinfo {pages}
  {235301} (\bibinfo {year} {2018})}\BibitemShut {NoStop}%
\bibitem [{\citenamefont {Ferioli}\ \emph {et~al.}(2019)\citenamefont
  {Ferioli}, \citenamefont {Semeghini}, \citenamefont {Masi}, \citenamefont
  {Giusti}, \citenamefont {Modugno}, \citenamefont {Inguscio}, \citenamefont
  {Gallem\'{\i}}, \citenamefont {Recati},\ and\ \citenamefont
  {Fattori}}]{Ferioli2019}%
  \BibitemOpen
  \bibfield  {author} {\bibinfo {author} {\bibfnamefont {G.}~\bibnamefont
  {Ferioli}}, \bibinfo {author} {\bibfnamefont {G.}~\bibnamefont {Semeghini}},
  \bibinfo {author} {\bibfnamefont {L.}~\bibnamefont {Masi}}, \bibinfo {author}
  {\bibfnamefont {G.}~\bibnamefont {Giusti}}, \bibinfo {author} {\bibfnamefont
  {G.}~\bibnamefont {Modugno}}, \bibinfo {author} {\bibfnamefont
  {M.}~\bibnamefont {Inguscio}}, \bibinfo {author} {\bibfnamefont
  {A.}~\bibnamefont {Gallem\'{\i}}}, \bibinfo {author} {\bibfnamefont
  {A.}~\bibnamefont {Recati}}, \ and\ \bibinfo {author} {\bibfnamefont
  {M.}~\bibnamefont {Fattori}},\ }\bibfield  {title} {\enquote {\bibinfo
  {title} {Collisions of self-bound quantum droplets},}\ }\href {\doibase
  10.1103/PhysRevLett.122.090401} {\bibfield  {journal} {\bibinfo  {journal}
  {Phys. Rev. Lett.}\ }\textbf {\bibinfo {volume} {122}},\ \bibinfo {pages}
  {090401} (\bibinfo {year} {2019})}\BibitemShut {NoStop}%
\bibitem [{\citenamefont {D'Errico}\ \emph {et~al.}(2019)\citenamefont
  {D'Errico}, \citenamefont {Burchianti}, \citenamefont {Prevedelli},
  \citenamefont {Salasnich}, \citenamefont {Ancilotto}, \citenamefont
  {Modugno}, \citenamefont {Minardi},\ and\ \citenamefont
  {Fort}}]{DErrico2019}%
  \BibitemOpen
  \bibfield  {author} {\bibinfo {author} {\bibfnamefont {C.}~\bibnamefont
  {D'Errico}}, \bibinfo {author} {\bibfnamefont {A.}~\bibnamefont
  {Burchianti}}, \bibinfo {author} {\bibfnamefont {M.}~\bibnamefont
  {Prevedelli}}, \bibinfo {author} {\bibfnamefont {L.}~\bibnamefont
  {Salasnich}}, \bibinfo {author} {\bibfnamefont {F.}~\bibnamefont
  {Ancilotto}}, \bibinfo {author} {\bibfnamefont {M.}~\bibnamefont {Modugno}},
  \bibinfo {author} {\bibfnamefont {F.}~\bibnamefont {Minardi}}, \ and\
  \bibinfo {author} {\bibfnamefont {C.}~\bibnamefont {Fort}},\ }\bibfield
  {title} {\enquote {\bibinfo {title} {Observation of quantum droplets in a
  heteronuclear bosonic mixture},}\ }\href {\doibase
  10.1103/PhysRevResearch.1.033155} {\bibfield  {journal} {\bibinfo  {journal}
  {Phys. Rev. Research}\ }\textbf {\bibinfo {volume} {1}},\ \bibinfo {pages}
  {033155} (\bibinfo {year} {2019})}\BibitemShut {NoStop}%
\bibitem [{\citenamefont {Rakshit}\ \emph
  {et~al.}(2019{\natexlab{a}})\citenamefont {Rakshit}, \citenamefont {Karpiuk},
  \citenamefont {Brewczyk},\ and\ \citenamefont {Gajda}}]{Rakshit2019}%
  \BibitemOpen
  \bibfield  {author} {\bibinfo {author} {\bibfnamefont {D.}~\bibnamefont
  {Rakshit}}, \bibinfo {author} {\bibfnamefont {T.}~\bibnamefont {Karpiuk}},
  \bibinfo {author} {\bibfnamefont {M.}~\bibnamefont {Brewczyk}}, \ and\
  \bibinfo {author} {\bibfnamefont {M.}~\bibnamefont {Gajda}},\ }\bibfield
  {title} {\enquote {\bibinfo {title} {Quantum {Bose}-{Fermi} droplets},}\
  }\href {\doibase 10.21468/SciPostPhys.6.6.079} {\bibfield  {journal}
  {\bibinfo  {journal} {SciPost Phys.}\ }\textbf {\bibinfo {volume} {6}},\
  \bibinfo {pages} {079} (\bibinfo {year} {2019}{\natexlab{a}})}\BibitemShut
  {NoStop}%
\bibitem [{\citenamefont {Rakshit}\ \emph
  {et~al.}(2019{\natexlab{b}})\citenamefont {Rakshit}, \citenamefont {Karpiuk},
  \citenamefont {Zin}, \citenamefont {Brewczyk}, \citenamefont {Lewenstein},\
  and\ \citenamefont {Gajda}}]{Rakshit2019a}%
  \BibitemOpen
  \bibfield  {author} {\bibinfo {author} {\bibfnamefont {D.}~\bibnamefont
  {Rakshit}}, \bibinfo {author} {\bibfnamefont {T.}~\bibnamefont {Karpiuk}},
  \bibinfo {author} {\bibfnamefont {P.}~\bibnamefont {Zin}}, \bibinfo {author}
  {\bibfnamefont {M.}~\bibnamefont {Brewczyk}}, \bibinfo {author}
  {\bibfnamefont {M.}~\bibnamefont {Lewenstein}}, \ and\ \bibinfo {author}
  {\bibfnamefont {M.}~\bibnamefont {Gajda}},\ }\bibfield  {title} {\enquote
  {\bibinfo {title} {Self-bound {Bose}-{Fermi} liquids in lower dimensions},}\
  }\href {\doibase 10.1088/1367-2630/ab2ce3} {\bibfield  {journal} {\bibinfo
  {journal} {New J. Phys.}\ }\textbf {\bibinfo {volume} {21}},\ \bibinfo
  {pages} {073027} (\bibinfo {year} {2019}{\natexlab{b}})}\BibitemShut
  {NoStop}%
\bibitem [{\citenamefont {Ferrier-Barbut}\ \emph {et~al.}(2016)\citenamefont
  {Ferrier-Barbut}, \citenamefont {Kadau}, \citenamefont {Schmitt},
  \citenamefont {Wenzel},\ and\ \citenamefont {Pfau}}]{Ferrier-Barbut2016}%
  \BibitemOpen
  \bibfield  {author} {\bibinfo {author} {\bibfnamefont {I.}~\bibnamefont
  {Ferrier-Barbut}}, \bibinfo {author} {\bibfnamefont {H.}~\bibnamefont
  {Kadau}}, \bibinfo {author} {\bibfnamefont {M.}~\bibnamefont {Schmitt}},
  \bibinfo {author} {\bibfnamefont {M.}~\bibnamefont {Wenzel}}, \ and\ \bibinfo
  {author} {\bibfnamefont {T.}~\bibnamefont {Pfau}},\ }\bibfield  {title}
  {\enquote {\bibinfo {title} {Observation of quantum droplets in a strongly
  dipolar {Bose} gas},}\ }\href {\doibase 10.1103/PhysRevLett.116.215301}
  {\bibfield  {journal} {\bibinfo  {journal} {Phys. Rev. Lett.}\ }\textbf
  {\bibinfo {volume} {116}},\ \bibinfo {pages} {215301} (\bibinfo {year}
  {2016})}\BibitemShut {NoStop}%
\bibitem [{\citenamefont {Chomaz}\ \emph {et~al.}(2016)\citenamefont {Chomaz},
  \citenamefont {Baier}, \citenamefont {Petter}, \citenamefont {Mark},
  \citenamefont {W\"achtler}, \citenamefont {Santos},\ and\ \citenamefont
  {Ferlaino}}]{Chomaz2016}%
  \BibitemOpen
  \bibfield  {author} {\bibinfo {author} {\bibfnamefont {L.}~\bibnamefont
  {Chomaz}}, \bibinfo {author} {\bibfnamefont {S.}~\bibnamefont {Baier}},
  \bibinfo {author} {\bibfnamefont {D.}~\bibnamefont {Petter}}, \bibinfo
  {author} {\bibfnamefont {M.~J.}\ \bibnamefont {Mark}}, \bibinfo {author}
  {\bibfnamefont {F.}~\bibnamefont {W\"achtler}}, \bibinfo {author}
  {\bibfnamefont {L.}~\bibnamefont {Santos}}, \ and\ \bibinfo {author}
  {\bibfnamefont {F.}~\bibnamefont {Ferlaino}},\ }\bibfield  {title} {\enquote
  {\bibinfo {title} {Quantum-fluctuation-driven crossover from a dilute
  {Bose-Einstein} condensate to a macrodroplet in a dipolar quantum fluid},}\
  }\href {\doibase 10.1103/PhysRevX.6.041039} {\bibfield  {journal} {\bibinfo
  {journal} {Phys. Rev. X}\ }\textbf {\bibinfo {volume} {6}},\ \bibinfo {pages}
  {041039} (\bibinfo {year} {2016})}\BibitemShut {NoStop}%
\bibitem [{\citenamefont {Ferrier-Barbut}\ \emph {et~al.}(2018)\citenamefont
  {Ferrier-Barbut}, \citenamefont {Wenzel}, \citenamefont {B\"ottcher},
  \citenamefont {Langen}, \citenamefont {Isoard}, \citenamefont {Stringari},\
  and\ \citenamefont {Pfau}}]{Ferrier-Barbut2018}%
  \BibitemOpen
  \bibfield  {author} {\bibinfo {author} {\bibfnamefont {I.}~\bibnamefont
  {Ferrier-Barbut}}, \bibinfo {author} {\bibfnamefont {M.}~\bibnamefont
  {Wenzel}}, \bibinfo {author} {\bibfnamefont {F.}~\bibnamefont {B\"ottcher}},
  \bibinfo {author} {\bibfnamefont {T.}~\bibnamefont {Langen}}, \bibinfo
  {author} {\bibfnamefont {M.}~\bibnamefont {Isoard}}, \bibinfo {author}
  {\bibfnamefont {S.}~\bibnamefont {Stringari}}, \ and\ \bibinfo {author}
  {\bibfnamefont {T.}~\bibnamefont {Pfau}},\ }\bibfield  {title} {\enquote
  {\bibinfo {title} {Scissors mode of dipolar quantum droplets of dysprosium
  atoms},}\ }\href {\doibase 10.1103/PhysRevLett.120.160402} {\bibfield
  {journal} {\bibinfo  {journal} {Phys. Rev. Lett.}\ }\textbf {\bibinfo
  {volume} {120}},\ \bibinfo {pages} {160402} (\bibinfo {year}
  {2018})}\BibitemShut {NoStop}%
\bibitem [{\citenamefont {Tanzi}\ \emph
  {et~al.}(2019{\natexlab{a}})\citenamefont {Tanzi}, \citenamefont {Lucioni},
  \citenamefont {Fam\`a}, \citenamefont {Catani}, \citenamefont {Fioretti},
  \citenamefont {Gabbanini}, \citenamefont {Bisset}, \citenamefont {Santos},\
  and\ \citenamefont {Modugno}}]{Tanzi2019}%
  \BibitemOpen
  \bibfield  {author} {\bibinfo {author} {\bibfnamefont {L.}~\bibnamefont
  {Tanzi}}, \bibinfo {author} {\bibfnamefont {E.}~\bibnamefont {Lucioni}},
  \bibinfo {author} {\bibfnamefont {F.}~\bibnamefont {Fam\`a}}, \bibinfo
  {author} {\bibfnamefont {J.}~\bibnamefont {Catani}}, \bibinfo {author}
  {\bibfnamefont {A.}~\bibnamefont {Fioretti}}, \bibinfo {author}
  {\bibfnamefont {C.}~\bibnamefont {Gabbanini}}, \bibinfo {author}
  {\bibfnamefont {R.~N.}\ \bibnamefont {Bisset}}, \bibinfo {author}
  {\bibfnamefont {L.}~\bibnamefont {Santos}}, \ and\ \bibinfo {author}
  {\bibfnamefont {G.}~\bibnamefont {Modugno}},\ }\bibfield  {title} {\enquote
  {\bibinfo {title} {Observation of a dipolar quantum gas with metastable
  supersolid properties},}\ }\href {\doibase 10.1103/PhysRevLett.122.130405}
  {\bibfield  {journal} {\bibinfo  {journal} {Phys. Rev. Lett.}\ }\textbf
  {\bibinfo {volume} {122}},\ \bibinfo {pages} {130405} (\bibinfo {year}
  {2019}{\natexlab{a}})}\BibitemShut {NoStop}%
\bibitem [{\citenamefont {Chomaz}\ \emph {et~al.}(2019)\citenamefont {Chomaz},
  \citenamefont {Petter}, \citenamefont {Ilzh\"ofer}, \citenamefont {Natale},
  \citenamefont {Trautmann}, \citenamefont {Politi}, \citenamefont
  {Durastante}, \citenamefont {van Bijnen}, \citenamefont {Patscheider},
  \citenamefont {Sohmen}, \citenamefont {Mark},\ and\ \citenamefont
  {Ferlaino}}]{Chomaz2019}%
  \BibitemOpen
  \bibfield  {author} {\bibinfo {author} {\bibfnamefont {L.}~\bibnamefont
  {Chomaz}}, \bibinfo {author} {\bibfnamefont {D.}~\bibnamefont {Petter}},
  \bibinfo {author} {\bibfnamefont {P.}~\bibnamefont {Ilzh\"ofer}}, \bibinfo
  {author} {\bibfnamefont {G.}~\bibnamefont {Natale}}, \bibinfo {author}
  {\bibfnamefont {A.}~\bibnamefont {Trautmann}}, \bibinfo {author}
  {\bibfnamefont {C.}~\bibnamefont {Politi}}, \bibinfo {author} {\bibfnamefont
  {G.}~\bibnamefont {Durastante}}, \bibinfo {author} {\bibfnamefont {R.~M.~W.}\
  \bibnamefont {van Bijnen}}, \bibinfo {author} {\bibfnamefont
  {A.}~\bibnamefont {Patscheider}}, \bibinfo {author} {\bibfnamefont
  {M.}~\bibnamefont {Sohmen}}, \bibinfo {author} {\bibfnamefont {M.~J.}\
  \bibnamefont {Mark}}, \ and\ \bibinfo {author} {\bibfnamefont
  {F.}~\bibnamefont {Ferlaino}},\ }\bibfield  {title} {\enquote {\bibinfo
  {title} {Long-lived and transient supersolid behaviors in dipolar quantum
  gases},}\ }\href {\doibase 10.1103/PhysRevX.9.021012} {\bibfield  {journal}
  {\bibinfo  {journal} {Phys. Rev. X}\ }\textbf {\bibinfo {volume} {9}},\
  \bibinfo {pages} {021012} (\bibinfo {year} {2019})}\BibitemShut {NoStop}%
\bibitem [{\citenamefont {B\"ottcher}\ \emph {et~al.}(2019)\citenamefont
  {B\"ottcher}, \citenamefont {Schmidt}, \citenamefont {Wenzel}, \citenamefont
  {Hertkorn}, \citenamefont {Guo}, \citenamefont {Langen},\ and\ \citenamefont
  {Pfau}}]{Bottcher2019}%
  \BibitemOpen
  \bibfield  {author} {\bibinfo {author} {\bibfnamefont {F.}~\bibnamefont
  {B\"ottcher}}, \bibinfo {author} {\bibfnamefont {J.-N.}\ \bibnamefont
  {Schmidt}}, \bibinfo {author} {\bibfnamefont {M.}~\bibnamefont {Wenzel}},
  \bibinfo {author} {\bibfnamefont {J.}~\bibnamefont {Hertkorn}}, \bibinfo
  {author} {\bibfnamefont {M.}~\bibnamefont {Guo}}, \bibinfo {author}
  {\bibfnamefont {T.}~\bibnamefont {Langen}}, \ and\ \bibinfo {author}
  {\bibfnamefont {T.}~\bibnamefont {Pfau}},\ }\bibfield  {title} {\enquote
  {\bibinfo {title} {Transient supersolid properties in an array of dipolar
  quantum droplets},}\ }\href {\doibase 10.1103/PhysRevX.9.011051} {\bibfield
  {journal} {\bibinfo  {journal} {Phys. Rev. X}\ }\textbf {\bibinfo {volume}
  {9}},\ \bibinfo {pages} {011051} (\bibinfo {year} {2019})}\BibitemShut
  {NoStop}%
\bibitem [{\citenamefont {Tanzi}\ \emph
  {et~al.}(2019{\natexlab{b}})\citenamefont {Tanzi}, \citenamefont {Roccuzzo},
  \citenamefont {Lucioni}, \citenamefont {Fam\`a}, \citenamefont {Fioretti},
  \citenamefont {Gabbanini}, \citenamefont {Modugno}, \citenamefont {Recati},\
  and\ \citenamefont {Stringari}}]{Tanzi2019Modes}%
  \BibitemOpen
  \bibfield  {author} {\bibinfo {author} {\bibfnamefont {L.}~\bibnamefont
  {Tanzi}}, \bibinfo {author} {\bibfnamefont {S.~M.}\ \bibnamefont {Roccuzzo}},
  \bibinfo {author} {\bibfnamefont {E.}~\bibnamefont {Lucioni}}, \bibinfo
  {author} {\bibfnamefont {F.}~\bibnamefont {Fam\`a}}, \bibinfo {author}
  {\bibfnamefont {A.}~\bibnamefont {Fioretti}}, \bibinfo {author}
  {\bibfnamefont {C.}~\bibnamefont {Gabbanini}}, \bibinfo {author}
  {\bibfnamefont {G.}~\bibnamefont {Modugno}}, \bibinfo {author} {\bibfnamefont
  {A.}~\bibnamefont {Recati}}, \ and\ \bibinfo {author} {\bibfnamefont
  {S.}~\bibnamefont {Stringari}},\ }\bibfield  {title} {\enquote {\bibinfo
  {title} {Supersolid symmetry breaking from compressional oscillations in a
  dipolar quantum gas},}\ }\href {\doibase 10.1038/s41586-019-1568-6}
  {\bibfield  {journal} {\bibinfo  {journal} {Nature (London)}\ }\textbf
  {\bibinfo {volume} {574}},\ \bibinfo {pages} {382} (\bibinfo {year}
  {2019}{\natexlab{b}})}\BibitemShut {NoStop}%
\bibitem [{\citenamefont {Guo}\ \emph {et~al.}(2019)\citenamefont {Guo},
  \citenamefont {B\"ottcher}, \citenamefont {Hertkorn}, \citenamefont
  {Schmidt}, \citenamefont {Wenzel}, \citenamefont {B\"uchler}, \citenamefont
  {Langen},\ and\ \citenamefont {Pfau}}]{Guo2019}%
  \BibitemOpen
  \bibfield  {author} {\bibinfo {author} {\bibfnamefont {M.}~\bibnamefont
  {Guo}}, \bibinfo {author} {\bibfnamefont {F.}~\bibnamefont {B\"ottcher}},
  \bibinfo {author} {\bibfnamefont {J.}~\bibnamefont {Hertkorn}}, \bibinfo
  {author} {\bibfnamefont {J.-N.}\ \bibnamefont {Schmidt}}, \bibinfo {author}
  {\bibfnamefont {M.}~\bibnamefont {Wenzel}}, \bibinfo {author} {\bibfnamefont
  {H.-P.}\ \bibnamefont {B\"uchler}}, \bibinfo {author} {\bibfnamefont
  {T.}~\bibnamefont {Langen}}, \ and\ \bibinfo {author} {\bibfnamefont
  {T.}~\bibnamefont {Pfau}},\ }\bibfield  {title} {\enquote {\bibinfo {title}
  {The low-energy {Goldstone} mode in a trapped dipolar supersolid},}\ }\href
  {\doibase 10.1038/s41586-019-1569-5} {\bibfield  {journal} {\bibinfo
  {journal} {Nature (London)}\ }\textbf {\bibinfo {volume} {574}},\ \bibinfo
  {pages} {386} (\bibinfo {year} {2019})}\BibitemShut {NoStop}%
\bibitem [{\citenamefont {W\"achtler}\ and\ \citenamefont
  {Santos}(2016{\natexlab{a}})}]{Wachtler2016}%
  \BibitemOpen
  \bibfield  {author} {\bibinfo {author} {\bibfnamefont {F.}~\bibnamefont
  {W\"achtler}}\ and\ \bibinfo {author} {\bibfnamefont {L.}~\bibnamefont
  {Santos}},\ }\bibfield  {title} {\enquote {\bibinfo {title} {Quantum
  filaments in dipolar {Bose-Einstein} condensates},}\ }\href {\doibase
  10.1103/PhysRevA.93.061603} {\bibfield  {journal} {\bibinfo  {journal} {Phys.
  Rev. A}\ }\textbf {\bibinfo {volume} {93}},\ \bibinfo {pages} {061603}
  (\bibinfo {year} {2016}{\natexlab{a}})}\BibitemShut {NoStop}%
\bibitem [{\citenamefont {W\"achtler}\ and\ \citenamefont
  {Santos}(2016{\natexlab{b}})}]{Wachtler2016a}%
  \BibitemOpen
  \bibfield  {author} {\bibinfo {author} {\bibfnamefont {F.}~\bibnamefont
  {W\"achtler}}\ and\ \bibinfo {author} {\bibfnamefont {L.}~\bibnamefont
  {Santos}},\ }\bibfield  {title} {\enquote {\bibinfo {title} {Ground-state
  properties and elementary excitations of quantum droplets in dipolar
  {Bose-Einstein} condensates},}\ }\href {\doibase 10.1103/PhysRevA.94.043618}
  {\bibfield  {journal} {\bibinfo  {journal} {Phys. Rev. A}\ }\textbf {\bibinfo
  {volume} {94}},\ \bibinfo {pages} {043618} (\bibinfo {year}
  {2016}{\natexlab{b}})}\BibitemShut {NoStop}%
\bibitem [{\citenamefont {Bisset}\ \emph {et~al.}(2016)\citenamefont {Bisset},
  \citenamefont {Wilson}, \citenamefont {Baillie},\ and\ \citenamefont
  {Blakie}}]{Bisset2016}%
  \BibitemOpen
  \bibfield  {author} {\bibinfo {author} {\bibfnamefont {R.~N.}\ \bibnamefont
  {Bisset}}, \bibinfo {author} {\bibfnamefont {R.~M.}\ \bibnamefont {Wilson}},
  \bibinfo {author} {\bibfnamefont {D.}~\bibnamefont {Baillie}}, \ and\
  \bibinfo {author} {\bibfnamefont {P.~B.}\ \bibnamefont {Blakie}},\ }\bibfield
   {title} {\enquote {\bibinfo {title} {Ground-state phase diagram of a dipolar
  condensate with quantum fluctuations},}\ }\href {\doibase
  10.1103/PhysRevA.94.033619} {\bibfield  {journal} {\bibinfo  {journal} {Phys.
  Rev. A}\ }\textbf {\bibinfo {volume} {94}},\ \bibinfo {pages} {033619}
  (\bibinfo {year} {2016})}\BibitemShut {NoStop}%
\bibitem [{\citenamefont {Petrov}\ and\ \citenamefont
  {Astrakharchik}(2016)}]{Petrov2016}%
  \BibitemOpen
  \bibfield  {author} {\bibinfo {author} {\bibfnamefont {D.~S.}\ \bibnamefont
  {Petrov}}\ and\ \bibinfo {author} {\bibfnamefont {G.~E.}\ \bibnamefont
  {Astrakharchik}},\ }\bibfield  {title} {\enquote {\bibinfo {title}
  {Ultradilute low-dimensional liquids},}\ }\href {\doibase
  10.1103/PhysRevLett.117.100401} {\bibfield  {journal} {\bibinfo  {journal}
  {Phys. Rev. Lett.}\ }\textbf {\bibinfo {volume} {117}},\ \bibinfo {pages}
  {100401} (\bibinfo {year} {2016})}\BibitemShut {NoStop}%
\bibitem [{\citenamefont {Khaykovich}\ \emph {et~al.}(2002)\citenamefont
  {Khaykovich}, \citenamefont {Schreck}, \citenamefont {Ferrari}, \citenamefont
  {Bourdel}, \citenamefont {Cubizolles}, \citenamefont {Carr}, \citenamefont
  {Castin},\ and\ \citenamefont {Salomon}}]{Khaykovich2002}%
  \BibitemOpen
  \bibfield  {author} {\bibinfo {author} {\bibfnamefont {L.}~\bibnamefont
  {Khaykovich}}, \bibinfo {author} {\bibfnamefont {F.}~\bibnamefont {Schreck}},
  \bibinfo {author} {\bibfnamefont {G.}~\bibnamefont {Ferrari}}, \bibinfo
  {author} {\bibfnamefont {T.}~\bibnamefont {Bourdel}}, \bibinfo {author}
  {\bibfnamefont {J.}~\bibnamefont {Cubizolles}}, \bibinfo {author}
  {\bibfnamefont {L.~D.}\ \bibnamefont {Carr}}, \bibinfo {author}
  {\bibfnamefont {Y.}~\bibnamefont {Castin}}, \ and\ \bibinfo {author}
  {\bibfnamefont {C.}~\bibnamefont {Salomon}},\ }\bibfield  {title} {\enquote
  {\bibinfo {title} {Formation of a matter-wave bright soliton},}\ }\href
  {\doibase 10.1126/science.1071021} {\bibfield  {journal} {\bibinfo  {journal}
  {Science}\ }\textbf {\bibinfo {volume} {296}},\ \bibinfo {pages} {1290}
  (\bibinfo {year} {2002})}\BibitemShut {NoStop}%
\bibitem [{\citenamefont {Strecker}\ \emph {et~al.}(2002)\citenamefont
  {Strecker}, \citenamefont {Partridge}, \citenamefont {Truscott},\ and\
  \citenamefont {Hulet}}]{Strecker2002}%
  \BibitemOpen
  \bibfield  {author} {\bibinfo {author} {\bibfnamefont {K.~E.}\ \bibnamefont
  {Strecker}}, \bibinfo {author} {\bibfnamefont {G.~B.}\ \bibnamefont
  {Partridge}}, \bibinfo {author} {\bibfnamefont {A.~G.}\ \bibnamefont
  {Truscott}}, \ and\ \bibinfo {author} {\bibfnamefont {R.~G.}\ \bibnamefont
  {Hulet}},\ }\bibfield  {title} {\enquote {\bibinfo {title} {Formation and
  propagation of matter-wave soliton trains},}\ }\href {\doibase
  10.1038/nature747} {\bibfield  {journal} {\bibinfo  {journal} {Nature
  (London)}\ }\textbf {\bibinfo {volume} {417}},\ \bibinfo {pages} {150}
  (\bibinfo {year} {2002})}\BibitemShut {NoStop}%
\bibitem [{\citenamefont {Marchant}\ \emph {et~al.}(2013)\citenamefont
  {Marchant}, \citenamefont {Billam}, \citenamefont {Wiles}, \citenamefont
  {Yu}, \citenamefont {Gardiner},\ and\ \citenamefont
  {Cornish}}]{Marchant2013}%
  \BibitemOpen
  \bibfield  {author} {\bibinfo {author} {\bibfnamefont {A.~L.}\ \bibnamefont
  {Marchant}}, \bibinfo {author} {\bibfnamefont {T.~P.}\ \bibnamefont
  {Billam}}, \bibinfo {author} {\bibfnamefont {T.~P.}\ \bibnamefont {Wiles}},
  \bibinfo {author} {\bibfnamefont {M.~M.~H.}\ \bibnamefont {Yu}}, \bibinfo
  {author} {\bibfnamefont {S.~A.}\ \bibnamefont {Gardiner}}, \ and\ \bibinfo
  {author} {\bibfnamefont {S.~L.}\ \bibnamefont {Cornish}},\ }\bibfield
  {title} {\enquote {\bibinfo {title} {Controlled formation and reflection of a
  bright solitary matter-wave},}\ }\href {\doibase 10.1038/ncomms2893}
  {\bibfield  {journal} {\bibinfo  {journal} {Nat. Commun.}\ }\textbf {\bibinfo
  {volume} {4}},\ \bibinfo {pages} {1865} (\bibinfo {year} {2013})}\BibitemShut
  {NoStop}%
\bibitem [{\citenamefont {Medley}\ \emph {et~al.}(2014)\citenamefont {Medley},
  \citenamefont {Minar}, \citenamefont {Cizek}, \citenamefont {Berryrieser},\
  and\ \citenamefont {Kasevich}}]{Medley2014}%
  \BibitemOpen
  \bibfield  {author} {\bibinfo {author} {\bibfnamefont {P.}~\bibnamefont
  {Medley}}, \bibinfo {author} {\bibfnamefont {M.~A.}\ \bibnamefont {Minar}},
  \bibinfo {author} {\bibfnamefont {N.~C.}\ \bibnamefont {Cizek}}, \bibinfo
  {author} {\bibfnamefont {D.}~\bibnamefont {Berryrieser}}, \ and\ \bibinfo
  {author} {\bibfnamefont {M.~A.}\ \bibnamefont {Kasevich}},\ }\bibfield
  {title} {\enquote {\bibinfo {title} {Evaporative production of bright atomic
  solitons},}\ }\href {\doibase 10.1103/PhysRevLett.112.060401} {\bibfield
  {journal} {\bibinfo  {journal} {Phys. Rev. Lett.}\ }\textbf {\bibinfo
  {volume} {112}},\ \bibinfo {pages} {060401} (\bibinfo {year}
  {2014})}\BibitemShut {NoStop}%
\bibitem [{\citenamefont {Nguyen}\ \emph {et~al.}(2014)\citenamefont {Nguyen},
  \citenamefont {Dyke}, \citenamefont {Luo}, \citenamefont {Malomed},\ and\
  \citenamefont {Hulet}}]{Nguyen2014}%
  \BibitemOpen
  \bibfield  {author} {\bibinfo {author} {\bibfnamefont {J.~H.~V.}\
  \bibnamefont {Nguyen}}, \bibinfo {author} {\bibfnamefont {P.}~\bibnamefont
  {Dyke}}, \bibinfo {author} {\bibfnamefont {D.}~\bibnamefont {Luo}}, \bibinfo
  {author} {\bibfnamefont {B.}~\bibnamefont {Malomed}}, \ and\ \bibinfo
  {author} {\bibfnamefont {R.~G.}\ \bibnamefont {Hulet}},\ }\bibfield  {title}
  {\enquote {\bibinfo {title} {Collisions of matter-wave solitons},}\ }\href
  {\doibase 10.1038/nphys3135} {\bibfield  {journal} {\bibinfo  {journal} {Nat.
  Phys.}\ }\textbf {\bibinfo {volume} {10}},\ \bibinfo {pages} {918} (\bibinfo
  {year} {2014})}\BibitemShut {NoStop}%
\bibitem [{\citenamefont {Lepoutre}\ \emph {et~al.}(2016)\citenamefont
  {Lepoutre}, \citenamefont {Fouch\'e}, \citenamefont {Boiss\'e}, \citenamefont
  {Berthet}, \citenamefont {Salomon}, \citenamefont {Aspect},\ and\
  \citenamefont {Bourdel}}]{Lepoutre2016}%
  \BibitemOpen
  \bibfield  {author} {\bibinfo {author} {\bibfnamefont {S.}~\bibnamefont
  {Lepoutre}}, \bibinfo {author} {\bibfnamefont {L.}~\bibnamefont {Fouch\'e}},
  \bibinfo {author} {\bibfnamefont {A.}~\bibnamefont {Boiss\'e}}, \bibinfo
  {author} {\bibfnamefont {G.}~\bibnamefont {Berthet}}, \bibinfo {author}
  {\bibfnamefont {G.}~\bibnamefont {Salomon}}, \bibinfo {author} {\bibfnamefont
  {A.}~\bibnamefont {Aspect}}, \ and\ \bibinfo {author} {\bibfnamefont
  {T.}~\bibnamefont {Bourdel}},\ }\bibfield  {title} {\enquote {\bibinfo
  {title} {Production of strongly bound $^{39}\mathrm{K}$ bright solitons},}\
  }\href {\doibase 10.1103/PhysRevA.94.053626} {\bibfield  {journal} {\bibinfo
  {journal} {Phys. Rev. A}\ }\textbf {\bibinfo {volume} {94}},\ \bibinfo
  {pages} {053626} (\bibinfo {year} {2016})}\BibitemShut {NoStop}%
\bibitem [{\citenamefont {Di~Carli}\ \emph {et~al.}(2019)\citenamefont
  {Di~Carli}, \citenamefont {Colquhoun}, \citenamefont {Henderson},
  \citenamefont {Flannigan}, \citenamefont {Oppo}, \citenamefont {Daley},
  \citenamefont {Kuhr},\ and\ \citenamefont {Haller}}]{DiCarli2019}%
  \BibitemOpen
  \bibfield  {author} {\bibinfo {author} {\bibfnamefont {A.}~\bibnamefont
  {Di~Carli}}, \bibinfo {author} {\bibfnamefont {C.~D.}\ \bibnamefont
  {Colquhoun}}, \bibinfo {author} {\bibfnamefont {G.}~\bibnamefont
  {Henderson}}, \bibinfo {author} {\bibfnamefont {S.}~\bibnamefont
  {Flannigan}}, \bibinfo {author} {\bibfnamefont {G.-L.}\ \bibnamefont {Oppo}},
  \bibinfo {author} {\bibfnamefont {A.~J.}\ \bibnamefont {Daley}}, \bibinfo
  {author} {\bibfnamefont {S.}~\bibnamefont {Kuhr}}, \ and\ \bibinfo {author}
  {\bibfnamefont {E.}~\bibnamefont {Haller}},\ }\bibfield  {title} {\enquote
  {\bibinfo {title} {Excitation modes of bright matter-wave solitons},}\ }\href
  {\doibase 10.1103/PhysRevLett.123.123602} {\bibfield  {journal} {\bibinfo
  {journal} {Phys. Rev. Lett.}\ }\textbf {\bibinfo {volume} {123}},\ \bibinfo
  {pages} {123602} (\bibinfo {year} {2019})}\BibitemShut {NoStop}%
\bibitem [{\citenamefont {Hayata}\ and\ \citenamefont
  {Koshiba}(1995)}]{Hayata1995}%
  \BibitemOpen
  \bibfield  {author} {\bibinfo {author} {\bibfnamefont {K.}~\bibnamefont
  {Hayata}}\ and\ \bibinfo {author} {\bibfnamefont {M.}~\bibnamefont
  {Koshiba}},\ }\bibfield  {title} {\enquote {\bibinfo {title} {Algebraic
  solitary-wave solutions of a nonlinear schr\"odinger equation},}\ }\href
  {\doibase 10.1103/PhysRevE.51.1499} {\bibfield  {journal} {\bibinfo
  {journal} {Phys. Rev. E}\ }\textbf {\bibinfo {volume} {51}},\ \bibinfo
  {pages} {1499--1502} (\bibinfo {year} {1995})}\BibitemShut {NoStop}%
\bibitem [{\citenamefont {Triki}\ \emph {et~al.}(2017)\citenamefont {Triki},
  \citenamefont {Biswas}, \citenamefont {Moshokoa},\ and\ \citenamefont
  {Belic}}]{Triki2017}%
  \BibitemOpen
  \bibfield  {author} {\bibinfo {author} {\bibfnamefont {H.}~\bibnamefont
  {Triki}}, \bibinfo {author} {\bibfnamefont {A.}~\bibnamefont {Biswas}},
  \bibinfo {author} {\bibfnamefont {S.~P.}\ \bibnamefont {Moshokoa}}, \ and\
  \bibinfo {author} {\bibfnamefont {M.}~\bibnamefont {Belic}},\ }\bibfield
  {title} {\enquote {\bibinfo {title} {Optical solitons and conservation laws
  with quadratic-cubic nonlinearity},}\ }\href {\doibase
  https://doi.org/10.1016/j.ijleo.2016.10.010} {\bibfield  {journal} {\bibinfo
  {journal} {Optik}\ }\textbf {\bibinfo {volume} {128}},\ \bibinfo {pages} {63
  -- 70} (\bibinfo {year} {2017})}\BibitemShut {NoStop}%
\bibitem [{\citenamefont {Mithun}\ \emph {et~al.}(2020)\citenamefont {Mithun},
  \citenamefont {Maluckov}, \citenamefont {Kasamatsu}, \citenamefont
  {Malomed},\ and\ \citenamefont {Khare}}]{Mithun2020}%
  \BibitemOpen
  \bibfield  {author} {\bibinfo {author} {\bibfnamefont {T.}~\bibnamefont
  {Mithun}}, \bibinfo {author} {\bibfnamefont {A.}~\bibnamefont {Maluckov}},
  \bibinfo {author} {\bibfnamefont {K.}~\bibnamefont {Kasamatsu}}, \bibinfo
  {author} {\bibfnamefont {B.~A.}\ \bibnamefont {Malomed}}, \ and\ \bibinfo
  {author} {\bibfnamefont {A.}~\bibnamefont {Khare}},\ }\bibfield  {title}
  {\enquote {\bibinfo {title} {Modulational instability, inter-component
  asymmetry, and formation of quantum droplets in one-dimensional binary {Bose}
  gases},}\ }\href {\doibase 10.3390/sym12010174} {\bibfield  {journal}
  {\bibinfo  {journal} {Symmetry}\ }\textbf {\bibinfo {volume} {12}},\ \bibinfo
  {pages} {174} (\bibinfo {year} {2020})}\BibitemShut {NoStop}%
\bibitem [{\citenamefont {Astrakharchik}\ and\ \citenamefont
  {Malomed}(2018)}]{Astrakharchik2018}%
  \BibitemOpen
  \bibfield  {author} {\bibinfo {author} {\bibfnamefont {G.~E.}\ \bibnamefont
  {Astrakharchik}}\ and\ \bibinfo {author} {\bibfnamefont {B.~A.}\ \bibnamefont
  {Malomed}},\ }\bibfield  {title} {\enquote {\bibinfo {title} {Dynamics of
  one-dimensional quantum droplets},}\ }\href {\doibase
  10.1103/PhysRevA.98.013631} {\bibfield  {journal} {\bibinfo  {journal} {Phys.
  Rev. A}\ }\textbf {\bibinfo {volume} {98}},\ \bibinfo {pages} {013631}
  (\bibinfo {year} {2018})}\BibitemShut {NoStop}%
\bibitem [{\citenamefont {Kaup}(1990)}]{Kaup1990}%
  \BibitemOpen
  \bibfield  {author} {\bibinfo {author} {\bibfnamefont {D.~J.}\ \bibnamefont
  {Kaup}},\ }\bibfield  {title} {\enquote {\bibinfo {title} {Perturbation
  theory for solitons in optical fibers},}\ }\href {\doibase
  10.1103/PhysRevA.42.5689} {\bibfield  {journal} {\bibinfo  {journal} {Phys.
  Rev. A}\ }\textbf {\bibinfo {volume} {42}},\ \bibinfo {pages} {5689--5694}
  (\bibinfo {year} {1990})}\BibitemShut {NoStop}%
\bibitem [{\citenamefont {Pelinovsky}\ \emph {et~al.}(1998)\citenamefont
  {Pelinovsky}, \citenamefont {Kivshar},\ and\ \citenamefont
  {Afanasjev}}]{Pelinovsky1998}%
  \BibitemOpen
  \bibfield  {author} {\bibinfo {author} {\bibfnamefont {D.~E.}\ \bibnamefont
  {Pelinovsky}}, \bibinfo {author} {\bibfnamefont {Yu.~S.}\ \bibnamefont
  {Kivshar}}, \ and\ \bibinfo {author} {\bibfnamefont {V.~V.}\ \bibnamefont
  {Afanasjev}},\ }\bibfield  {title} {\enquote {\bibinfo {title} {Internal
  modes of envelope solitons},}\ }\href {\doibase
  10.1016/S0167-2789(98)80010-9} {\bibfield  {journal} {\bibinfo  {journal}
  {Physica D}\ }\textbf {\bibinfo {volume} {116}},\ \bibinfo {pages} {121}
  (\bibinfo {year} {1998})}\BibitemShut {NoStop}%
\bibitem [{\citenamefont {Birnbaum}\ and\ \citenamefont {Malomed}(2008)}]{Birnbaum2008}%
  \BibitemOpen
  \bibfield  {author} {\bibinfo {author} {\bibfnamefont {Z.}\ \bibnamefont
  {Birnbaum}}\ and\ \bibinfo {author} {\bibfnamefont {B.~A.}~\bibnamefont {Malomed}},\
  }\bibfield  {title} {\enquote {\bibinfo {title} {Families of spatial solitons
in a two-channel waveguide with the cubic-quintic nonlinearity},}\ }\href {\doibase 
  10.1016/j.physd.2008.08.005} {\bibfield  {journal} {\bibinfo  {journal} {Physica D}\ }\textbf {\bibinfo {volume} {237}},\ \bibinfo {pages} {3252}
  (\bibinfo {year} {2008})}\BibitemShut {NoStop}%
\bibitem [{\citenamefont {Sinha}\ \emph {et~al.}(2006)\citenamefont {Sinha},
  \citenamefont {Cherny}, \citenamefont {Kovrizhin},\ and\ \citenamefont
  {Brand}}]{Sinha2006}%
  \BibitemOpen
  \bibfield  {author} {\bibinfo {author} {\bibfnamefont {S.}~\bibnamefont
  {Sinha}}, \bibinfo {author} {\bibfnamefont {A.~Yu.}\ \bibnamefont {Cherny}},
  \bibinfo {author} {\bibfnamefont {D.}~\bibnamefont {Kovrizhin}}, \ and\
  \bibinfo {author} {\bibfnamefont {J.}~\bibnamefont {Brand}},\ }\bibfield
  {title} {\enquote {\bibinfo {title} {Friction and diffusion of matter-wave
  bright solitons},}\ }\href {\doibase 10.1103/PhysRevLett.96.030406}
  {\bibfield  {journal} {\bibinfo  {journal} {Phys. Rev. Lett.}\ }\textbf
  {\bibinfo {volume} {96}},\ \bibinfo {pages} {030406} (\bibinfo {year}
  {2006})}\BibitemShut {NoStop}%
\bibitem [{\citenamefont {Carr}\ and\ \citenamefont {Castin}(2002)}]{Carr2002}%
  \BibitemOpen
  \bibfield  {author} {\bibinfo {author} {\bibfnamefont {L.~D.}\ \bibnamefont
  {Carr}}\ and\ \bibinfo {author} {\bibfnamefont {Y.}~\bibnamefont {Castin}},\
  }\bibfield  {title} {\enquote {\bibinfo {title} {Dynamics of a matter-wave
  bright soliton in an expulsive potential},}\ }\href {\doibase
  10.1103/PhysRevA.66.063602} {\bibfield  {journal} {\bibinfo  {journal} {Phys.
  Rev. A}\ }\textbf {\bibinfo {volume} {66}},\ \bibinfo {pages} {063602}
  (\bibinfo {year} {2002})}\BibitemShut {NoStop}%
\end{thebibliography}
\end{document}